# Econometric model of children participation in family dairy farming in the center of dairy farming, West Java Province, Indonesia

**Achmad FIRMAN[1*] and Ratna Ayu SAPTATI[2]**

[1]*Department of Social Economics, Animal Husbandry Faculty, Universitas Padjadjaran*
*Jl. Raya Bandung-Sumedang Km 21, Jatinangor, Sumedang 4365*
[2]*Indonesian Center for Animal Research and Development, Ministry of Agriculture*
*Jl. Raya Pajajaran Kav E-59, Bogor 16151*
*Corresponding author: achmad.firman@unpad.ac.id

**Abstract:** The involvement of children in the family dairy farming is pivotal point to reduce the cost of production input, especially in smallholder dairy farming. The purposes of the study are to analysis the factors that influence children's participation in working in the family dairy farm. The study was held December 2020 in the development center of dairy farming in Pangalengan subdistrict, West Java Province, Indonesia. The econometric method used in the study was the logit regression model. The results of the study determine that the there were number of respondents who participates in family farms was 52.59% of total respondents, and the rest was no participation in the family farms. There are 3 variables in the model that are very influential on children's participation in the family dairy farming, such as $X_1$ (number of dairy farm land ownership), $X_2$ (number of family members), and $X_6$ (the amount of work spent on the family's dairy farm).
**Key words:** Participation, children, family, dairy farming, logit model

## 1. Introduction

Dairy farming has played an important role in the developing countries, including Indonesia. The farms are dominated by smallholder farmers which rise three to five cows per farmer (Mandaka and Hutagaol, 2005 cited by Anindyasari et al 2015; DGLH 2011 cited by Daud et al 2015). However, smallholder farmers have a fairly large role for the food security in their families, economic sustainability, job creation, and social networking (FAO 2018). The features of smallholder dairy farmers have 1 to 5 cows, inadequate capital and assets, semi-permanent stalls, conventional management, family-based labor, poor and food insecure, having several economic activities, and insufficient money to raise dairy cows (Asmara et al 2017; Firman et al 2018; Rapsomanikis 2015).

Small-scale agricultural terminology describes the family farm business. In developing countries, farm business is operated by a family. Family farmers become the biggest source of employment around the world, contribute to the rural development and food security (FAO 2018; Mbah et all 2016). In general, family farming applies a multicultural farming system. This multiple agriculture activity provides an opportunity for family farmers to get income from several agricultural activities. For example, dairy farmers not only raise dairy cows, but also manage agricultural land. This is aimed at maintaining food source and family income. Therefore, this multiple agriculture activity is more likely to protect the family farm life from the failure of one agricultural activity.

The dairy business has become the main source of livelihood for most of the dairy farmers in rural and many dairy cattle development centers in Indonesia, one of them is in Pangalengan sub-district in Indonesia. This business is an integral part of the life of dairy farmers because it contributes to the family income. Smallholder farmers raise livestock with simple maintenance, which involves the entire family become an inseparable part of their life (Sudarmanto et al 2005) including their children. Thus, in a small-scale business, the involvement of each family member has an important role in reducing production costs. Generally, each family member has their respective duties in running their business, including children (Wisaptiningsih et al 2018).

By definition, child is someone who is not yet 18 (eighteen) years old, including children who are still in the womb (Law of the Republic of Indonesia Number 23 of 2002 concerning Child Protection, article 1 Paragraph 1). Meanwhile, according to the WHO definition, the age limit of a child is from the womb until 19 years old (Infodatin, 2014). Therefore, children who involve in the study is in the range of 7 – 19 years old.

According to FAO, Approoximately 70% of child labour work in agriculture sector and many children work in hazardous activities (FAO 2006). In contrast, children who involve in family dairy farming in Indonesia work on the basis of his will. They participate to help parents in the farming because the dairy farming is a part of their life. The participation of children in the family dairy cow business is more likely to take advantage of their time outside of school hours, such as looking for grass, cleaning cow stalls, washing cows, and feeding. According to Marina et al (2013), the portion of children helping the family's dairy cow business is only 10% of the total working time per day. Although the contribution is very low, children's participation can ease the workload of parents and reduce production costs. In addition, if the contribution is collected in a month, children's participation can reduce the burden of production costs, especially from labor costs. Therefore, economic factors are the main choice for involving family labor. In relation to neo-classical economic theory, it explains that every individual desire to maximize profits and will always try to find economic activities that benefit (Hennessey 2002). The contribution of this child is also able to lift the family out of poverty. Based on Satrio (2018), the agricultural sector makes a large contribution to poverty in West Java Province. Based on this, it can be presumed that the role of children in the family dairy cow business has added value in reducing family poverty.

Children are an important input factor in a household business. Children's participation has an important role in family farming. The use of child labor in household dairy farming, in addition to reducing costs, it also prepares job opportunities. The interaction process in managing family businesses will directly or indirectly result in the transfer of norms, culture, knowledge, behavior, and life lessons from parents to children (Fennell 1981 cited by Bohak and Borec 2009; Lobley 2010). This process is called the socialization process, which is the process of transferring habits or values and rules or culture from one generation to another in a family, group or society. When individuals are young, the role of the family is very large in disseminating norms, culture, values and rules in the family, including their business activities. However, when these individuals have begun to interact with communities outside the family, such as at school, their neighborhood, religious institution, and in other communities, the environment will also play a major role in influencing children's thinking in the future. The younger generation will acquire knowledge, norms, and values of the society out of the family. Therefore, the structural changes in society caused by increasing levels of education, income levels, media development, relationships, and other external factors are very likely to affect the young farmers (Crisogen 2015; Passo and Urbano 2017). Moreover, access to information technology, such as a mobile phone connected to the internet, can provide sufficiently broad information and change insights for users, including children. Yogaswara (2015) cited by Firman, et al (2018) said that the technological information and communication in various ways has influenced the younger generation's perception of working on farms. Children's participation in family businesses is part of the primary socialization process. According to Firman et al (2018) determined that primary socialization is important for the regeneration process because the involvement of children in the dairy farming can make a deep impression on them. Therefore, the objectives of this study were to know how many children, as respondents, participate or do not participate in the family dairy cow business, and the factors that influenced the participation of children in the family dairy business.

## 2. Materials and methods

The research was carried out from December 2020 at the center for developing a dairy farm in Pangalengan subdistrict, West Java Province, Indonesia. The research location was chosen based on the consideration that the location was the first-time dairy cows were introduced by outsiders. Respondents were children (both males and females) of farmers aged 7-19 years (school ages) who are still living in one house with their parents, haven't married yet, and only one child per household was interviewed.

**2.1. Data collection.** Data was collected from 116 farmers out of 154 dairy farmers who have children aged 7 – 19 years in 5 Milk Collection Point (MCP) in Pangalengan subdistrict which were selected based on proportional random sampling. Heterogeneity data testing is measured using the coefficient of variation (CV). Mathematically, CV is formulated from the division of the standard deviation by the mean value (Hendayana 2013; Setiawan 2012). In the study, the CV value is less than 10% indicating homogeneous data, therefore these variables must be excluded from the formula. The variables can be seen in Table 1. Based on the Table 1 can be shown that there were no independent variables that have CV value under 10%. Therefore, all variables can be included in the model.

**Table 1. The characteristic of variables and coefficient of variation**

| Variables | Definition | Stand.Dev | Average | CV (%) |
|---|---|---|---|---|
| $Y$ | Children's participation, Y=1; not participation, Y=0 | 0.50 | 0.53 | 95.37 |
| $X_1$ | Number of dairy farm land ownership ($m^2$) | 66.56 | 91.96 | 72.38 |
| $X_2$ | Number of family members | 0.89 | 3.69 | 24.09 |
| $X_3$ | Order of children in the family; $1^{st}$, $2^{nd}$, $3^{rd}$, and etc | 0.66 | 1.35 | 48.97 |
| $X_4$ | Age of the children | 2.28 | 15.25 | 14.97 |
| $X_5$ | Formal level education (elementary school = 1; junior high school = 2; senior high school = 3) | 0.61 | 2.38 | 25.79 |
| $X_6$ | The amount of work spent on the family's dairy farm (hours per day) | 2.50 | 2.66 | 94.25 |
| $X_7$ | Number of dairy cows ownership (heads) | 3.20 | 5.17 | 61.85 |

| Variables | Definition | Stand.Dev | Average | CV (%) |
|---|---|---|---|---|
| $X_8$ | Dairy cow productivity (liter of milk/head/day) | 3.15 | 12.51 | 25.16 |
| $D_1$ | Gender (male = 1; female = 0) | 0.48 | 0.65 | 74.35 |

**2.2. Models analysis.** The logit regression model is used in this study because there are two binary events, namely participation of children in family dairy farming and not participating. The similar model is used in the children's involvement in family business in Attica, Greece (Pastrapa et al 2017). In addition, logit models more precisely predict and interpret variable outcome (Nawangsih 2013). The logistic regression model is a technique of econometric analysis to represent the connection between the dependent variable which has two or more categories with one or more independent variable of the type or interval (Hosmer and Lemeshow, 2000). According to Agresti (2007), logistic regression is a nonlinear regression to describe the relationship between Y (dependent variables) and X (independent variables), where the occurrence of variable Y is binary.

The binary event in the study is the participation of children in family dairy farming (Y = 1) and not participating in family dairy farming (Y = 0). According to Mc Cullagh and Nelder (1989) cited by Hendayana (2013) and Agresti (2007) determine that the binary logis regression follows the Bernoulli distribution as follows:

$$f(Y_i \mid \pi(x_i) = \pi(x_i)^{y_i}(1 - \pi(x_i))^{1-y_i} \quad \text{...................(1)}$$

In the equation (1), the value $y_i$ is 0 or 1, and $\pi(x_i)$ is the variable possibility of the i incident. The $\pi(x_i)$ can be formulated as follows:

$$\pi(x_i) = \frac{e^{(\alpha + \sum_{j=1}^{n} \beta_i X_{ji} + \sum_{k=1}^{m} \gamma_k D_{ki})}}{1 + e^{(\alpha + \sum_{j=1}^{n} \beta_i X_{ji} + \sum_{k=1}^{m} \gamma_k D_{ki})}} \quad \text{...................(2)}$$

if :
$$z = \alpha + \sum_{j=1}^{n} \beta_i X_{ji} + \sum_{k=1}^{m} \gamma_k D_{ki}) \quad \text{...................(3)}$$

The equation 2 can made a simple formula as follows:

$$\pi(x_i) = \frac{e^z}{1 + e^z} \quad \text{...................(4)}$$

According to Hosmer and Lemeshow (2000), the transformation of $\pi(x_i)$ is the logit transformation. The transformation can be defined, in terms of $\pi(x_i)$, as:

$$g(x_i) = \ln \frac{\pi(x_i)}{(1 - \pi(x_i))} = z = \alpha + \sum_{j=1}^{n} \beta_i X_{ji} + \sum_{k=1}^{m} \gamma_k D_{ki}) + \varepsilon \quad \text{...................(5)}$$

In multiple linear regression, it is assumed that $y = \pi(x_i) + \varepsilon$, where $\varepsilon$ is the sources of data variations that cannot be included in the model (error) and shows the difference between the object of observation and the expected value. The quantity of $\varepsilon$ is assumed normally distributed with a constant zero variance of the observed variables and may have one of two possible values (Hosmer and Lemeshow, 2000; Tinuki, 2010). If y = 1 then $\varepsilon = 1 - \pi(x_i)$ with probability $\pi(x_i)$, and if y = 0 then $\varepsilon = - \pi(x_i)$ with probability $1 - \pi(x_i)$. Thus, $\varepsilon$ has a distribution mean zero to $\pi(x_i)[1 - \pi(x_i)]$. Based on equation 5, $\pi(x_i)$ is the participation of children in the family dairy farming (y = 1), $1 - \pi(x_i)$ is children do not participate in the family dairy farm(y = 0); $\frac{\pi(x_i)}{(1 - \pi(x_i))} =$ Odds ratio; $X_{ji}$ = Vector of independent variables (j = 1,2, …..n); $D_{ki}$ =Vector of dummy variables (k = 1,2, …. n); $\alpha, \beta_i, \gamma_k$ = parameters of alleged logistics function of random error.

The logit model must be tested in order to make the model is goodness of fit. There are several steps to test the model (Hendayana (2013) as follows:
a. G test is based on hypothesis. The rule of the decision is to refuse $H_0$ if $G_{count} > \chi^2_{\alpha(p)}$. If $H_0$ is refused, it means that model is significant ($\alpha$).
b. Wald test (W). The rule of the decision is to refuse $H_0$ if $|W_{count}| > Z_{\alpha/2}$. If $H_0$ is refused, these parameters are statistically significant.
c. Concordant, Discordant and Ties are used to measure of association between the response variable and predicted probabilities.

## 3. Results
### 3.1. Respondents profile

In the study location, smallholder dairy farmers who raise 1-6 cows (Firman 2018; Asmara et al 2017) dominate dairy farming. The condition has not changed since the 1980s, however the business has become their livelihood. The children of the dairy farming families are the spearhead of the family dairy farming in order to the family business remains sustainable. The profile of children from families of dairy farmers who became respondents can be seen in Table 2. The age of the children who became respondents was dominated by those aged 13-19 years. Generally, respondents are at school age range, therefore the age characterizes were similar to the period of formal education they are currently undergoing. Most respondents were the oldest child in their family with 1 to 5 family members. In addition, the respondents who were interviewed were generally males. Number of respondents who participates in family farms was 61 respondents (52.59%) and the rest was no participation in the family farms. No participation means that children did not involve or help their parents in the family dairy farms.

**Table 2. Identification of children from families of dairy farmers in the Centre of Dairy Farming in Pangalengan subdistrict**

| No | Identification | Criteria | % |
|---|---|---|---|
| 1 | Ages (years old) | 8 – 12 | 6.90 |
|   |   | 13 – 15 | 50.00 |
|   |   | 16 – 19 | 43.10 |
| 2 | Formal education (on going status) | Elementary school | 6.90 |
|   |   | Junior high school | 50.00 |
|   |   | Senior high school | 43.10 |
| 3 | Order of children in the family | 1st | 75.00 |
|   |   | 2nd | 14.66 |
|   |   | 3rd | 10.34 |

| No | Identification | Criteria | % |
|---|---|---|---|
| 4 | Family members (person) | 1-5 persons | 97.41 |
|   |   | >5 persons | 2.59 |
| 5 | Gender | Male | 64.66 |
|   |   | female | 35.34 |
| 6 | Number of respondents participating in family dairy farm | No participation | 47.41 |
|   |   | Participation | 52.59 |

Note: n = 116 respondents

**Factors influence children to participate or not to participate in the family dairy business**

The participation of children in helping the family dairy cow business has become part of their life in the study area. The participation of children in family dairy business is limited to a number of activities from all dairy farming activities, such as cleaning pens, feeding, or looking for grass because they have to share activities with schools or other activities outside of school. Therefore, this study aims to determine what factors encourage these children to participate or not participate in the dairy cow family business. The logit model is used to determine what factors have a major influence on these conditions. The logit model formula of children's involvement in family dairy farming is shown below. There were 9 independent variables (including one dummy variable) involved in the model and all variables could be involved in the model because the CV > 10%. The logit model of the study as follows:

$$Ln\frac{\pi i}{(1-\pi i)} = \alpha + \beta_1 X_1 + \beta_2 X_2 + \beta_3 X_3 + \beta_4 X_4 + \beta_5 X_5 + \beta_6 X_6 + \beta_7 X_7 + \beta_8 X_8 + \gamma_1 D_1 + \varepsilon \quad \ldots\ldots\ldots\ldots(6)$$

Following the model, $X_1$ = number of dairy farm land ownership, $X_2$ = Number of family members, $X_3$ = Order of children in the family, $X_4$ = Age of the children, $X_5$ = Formal level education, $X_6$ = The amount of work spent on the family's dairy farm, $X_7$ = Number of dairy cow ownership, $X_8$ = Number of dairy cow ownership, and $D_1$ = dummy Gender.

The formulation must be tested before it can be used as a valid model. Formulation testing was carried out in several stages, namely the Pearson Method, Deviance Method, Hosmer-Lemeshow Method, Overall Model Test with G Test, and Partial Test with Wald (W). The tests and results can be seen in Table 3. According to Table 3, the results obtained for the overall model test show the value of a Log-likelihood of -10.763 with G statistics of 138.973 and p-value 0.000. Because the P-value was far below in α = 0.05, it could be said that the overall design of the Logistic Regression Model of the study was good for the model. Goodness of Fit uses the Pearson, Deviance and Hosmer-Lemeshow methods gave Chi-square results count 44.852; 21.526 and 0.246 with their respective p-values 0.886; 0.865 and 0.763. The p-value was above the α = 0.05 so that $H_0$ was accepted. Based on Hendayana (2013), it shows that the p-value is above the value of α = 0.05, the logit model is goodness of fit.

**Table 3. The results of logit regression model**

| Variables | Coefficient | Stand.Dev | Z | P | Odds ratio |
|---|---|---|---|---|---|
| Constant | -33.9 | 747.4 | -0.05 | 0.964 | NA |
| $X_1$ | 0.012388 | 0.008043 | 1.54 | 0.023* | 1.01 |
| $X_2$ | -2.526 | 1.423 | -1.78 | 0.036* | 1.08 |
| $X_3$ | 1.241 | 1.273 | 0.97 | 0.830 | 3.46 |
| $X_4$ | 0.6616 | 0.6111 | 1.08 | 0.279 | 1.94 |
| $X_5$ | 1.070 | 2.922 | 0.37 | 0.714 | 2.91 |
| $X_6$ | 29.8 | 747.4 | 0.04 | 0.018* | 8.77 |
| $X_7$ | 0.4254 | 0.2858 | 1.49 | 0.137 | 1.53 |
| $X_8$ | -0.7503 | 0.4957 | -1.51 | 0.130 | 0.47 |
| $D_1$ | -1.117 | 1.417 | -0.79 | 0.431 | 0.33 |

Log-Likelihood = -10.763
Test that all slopes are zero: G = 138.973, DF = 9, P-Value = 0.000

Goodness-of-Fit Tests

| Method | Chi-Square | DF | P |
|---|---|---|---|
| Pearson | 44.852 | 106 | 0.886 |
| Deviance | 21.526 | 106 | 0.865 |
| Hosmer-Lemeshow | 0.246 | 8 | 0.763 |

Measures of Association:
(Between the Response Variable and Predicted Probabilities)

| Pairs | Number | Percent | Summary | Measures |
|---|---|---|---|---|
| Concordant | 3319 | 92.7% | Somers D | 0.98 |
| Discordant | 251 | 7.0% | Goodman-Kruskal Gamma | 0.99 |
| Ties | 11 | 0.3% | Kendall's Tau-a | 0.50 |
| Total | 3581 | 100.0% | | |

Model testing was also measured from the association measurement to measure the predictive power of the model. The predictive power of the model was measured from the Concordant, Discordant, and Ties values. The predicted value is better if the Concordant value is greater than the Discordant and Ties value. The results have shown that the Concordant, Discordant, and Ties values were 92.7%, 7.0% and 0.3%, respectively. The Concordant value of 92.7 percent means that about 92.7 percent of the observations were by category participate in the family dairy business (Y = 1) was thought to have more opportunities most of the children category do not participate in the family dairy business. The value of Discordant 7 percent observations that do not participate in the family dairy business (Y = 0) had a greate chance compared with children participating in the family dairy business. The Ties value of 0.3 percent shows the percentage of observations with the category of participating children was the same as those who do not participate in the family dairy cow business (could be ignored). Based on the test results from several test models above, it can be concluded that the Logistic Regression Model of the study was goodness of fit to be used in this study.

To look at partial test, the test used Wald (W) test that categoriezed by Z coefficient per each variable. According to Table 3, there were 3 independent variables that have p-value below α = 0.05 (a very significant effect), namely $X_1$ (number of dairy farm land ownership), $X_2$ (number of family members), and $X_6$ (the amount of work spent on the family's dairy farm). However, other explanatory variables have no significant effect in the model.

*Interpretation of results*
Number of dairy farm land ownership ($X_1$) was very influential on children's participation in the family dairy business because it had a confidence level of 95% as indicated by the Z test coefficient value of 1.54 with a p-value (0.023). The odds ratio coefficient contains 1.01. This means that children have more opportunities to participate in the family who have a large land ownership. Parents who have large agricultural land need children's participation to reduce the burden of production costs, especially labor outside the family.

The number of family members ($X_2$) was the independent variable which also had a strong influence on children's participation in the family dairy business. The variable had a confidence level of 95% as indicated by the Z test coefficient value of -1.78 with a p-value (0.036). The odds ratio coefficient contains 1.08. It means that children who had a higher number of family members tend to participate in the family dairy cow business, especially boys who are either the oldest or the 2nd or 3rd child. According to Mishra and Hisam (2007) and Firman (2019), participation of children on family farm can stimulate childeren's decision to help and involve in the family farms.

The amount of work spent on the family's dairy farm ($X_3$) was the independent variable which also had a strong influence on children's participation in the family dairy business. The variable had a confidence level of 95% as indicated by the Z test coefficient value of 0.04 with a p-value (0.018). The odds ratio coefficient contains 8.77. It means that children who had more time to work on family dairy farms tend to participate in family businesses. This was also evidenced by the greater odds ratio value than the variables $X_1$ and $X_2$.

Other variables such as $X_3$ = order of children in the family, $X_4$ = age of the children, $X_5$ = formal level of education, $X_7$ = number of dairy cows ownership, $X_8$ = number of dairy cows ownership, and $D_1$ = dummy gender had no significant effect on participation of children in the family dairy business.

## 4. Acknowledgment
This research did not receive any specific grant from funding agencies in the public, commercial, or not-for-profit sectors.

## 5. Conclusions
It can be concluded that there were number of respondents who participates in family farms was 52.59% of total respondents, and the rest was no participation in the family farms. Meanwhile, there are 3 variables in the model that are very influential on children's participation in the family dairy business, namely $X_1$ (number of dairy farm land ownership), $X_2$ (number of family members), and $X_6$ (the amount of work spent on the family's dairy farm).